\documentclass[twocolumn]{IEEEtran}
\usepackage{graphicx}

\newcommand{\bfx}{\mbox{\boldmath $x$}}

\newcommand{\bfy}{\mbox{\boldmath $y$}}\newcommand{\bfv}{\mbox{\boldmath $v$}}\newcommand{\bfu}{\mbox{\boldmath $u$}}
 \newcommand{\bfz}{\mbox{\boldmath $z$}}

\newcommand{\beq}{\begin{equation}}\newcommand{\eeq}{\end{equation}}
\newtheorem{theorem}{Theorem}\def\endofproof{\hfill\rule{6pt}{6pt}}
\newtheorem{exam}{{\bf Example}}\newenvironment{example}{\begin{exam} \rm }{\normalsize \end{exam}}
\usepackage{xfloat}\usepackage{cite} \usepackage{amssymb}\usepackage{amsmath} 
\begin{document}
\title{Properties and constructions of constrained codes for DNA-based data storage
\author{Kees A. Schouhamer Immink $\;$ and $\;$ Kui Cai}
\thanks{Kees A. Schouhamer Immink is with Turing Machines Inc, Willemskade 15b-d, 3016 DK Rotterdam, The Netherlands. E-mail: immink@turing-machines.com.}
\thanks{Kui Cai is with Singapore University of Technology and Design (SUTD), 8 Somapah Rd, 487372, Singapore. E-mail: cai\_kui@sutd.edu.sg.}\\ \thanks{This work is supported by Singapore Ministry of Education Academic Research Fund Tier 2 MOE2016-T2-2-054}
}
\maketitle
\begin{abstract} We describe properties and constructions of constraint-based codes for DNA-based data storage which account for the maximum repetition length and AT/GC~balance. We present algorithms for computing the number of sequences with maximum repetition length and AT/GC~balance constraint. We describe routines for translating binary runlength limited and/or balanced strings into DNA strands, and compute the efficiency of such routines. We show that the implementation of AT/GC-balanced codes is straightforward accomplished with binary balanced codes. We present codes that account for both the maximum repetition length and AT/GC~balance. We compute the redundancy difference between the binary and a fully fledged quaternary approach.  \end{abstract}

%
\section{Introduction}  The first large-scale archival DNA-based storage architecture was implemented by Church {\em et al.}~\cite{Ch5} in 2012. Blawat {\em et al.}~\cite{Bl7} described successful experiments for storing and retrieving data blocks of 22~Mbyte of digital data in synthetic DNA. Ehrlich and Zielinski~\cite{Erl} further explored the limits of storage capacity of DNA-based storage architectures.

Naturally occurring DNA consists of four types of {\em nucleotides}: adenine (A), cytosine (C), guanine (G), and thymine (T). A DNA strand (or oligonucleotides, or oligo in short) is a linear sequence of these four nucleotides that are composed by DNA synthesizers. Binary source, or user, data are translated into the four types of nucleotides, for example, by mapping two binary source into a single nucleotide (nt).

Strings of nucleotides should satisfy a few elementary conditions, called {\em constraints}, in order to be less error prone. Repetitions of the same nucleotide, a {\em homopolymer run}, significantly increase the chance of sequencing errors~\cite{Bor},~\cite{Ros}, so that such long runs should be avoided. For example,
in~\cite{Ros}, experimental studies show that once the homopolymer run is larger than 4~nt, the sequencing error rate starts increasing significantly. In addition,~\cite{Ros} also reports that oligos with large unbalance between GC and AT content exhibit high dropout rates and are prone to polymerase chain reaction (PCR) errors, and should therefore be avoided.

Blawat's format~\cite{Bl7} incorporates a constrained code that uses a look-up table for translating binary source data into strands of nucleotides with a homopolymer run of length at most three. Blawat's format did not incorporate an AT/GC~balance constraint. Strands that do not satisfy the maximum homopolymer run requirement or the weak balance constraint are barred in Erlich's coding format~\cite{Erl}.

In this paper, we describe properties and constructions of quaternary constraint-based codes for DNA-based storage which account for a maximum homopolymer run and maximum unbalance between AT and GC contents. Binary `balanced' and runlength limited sequences have found widespread use in data communication and storage practice~\cite{Cat}. We show that constrained binary sequences can easily be translated into constrained quaternary sequences, which opens the door to a wealth of efficient binary code constructions for application in DNA-based storage~\cite{I72},~\cite{Ki3},~\cite{Che}. A further advantage of this binary approach instead of a `direct' 4-ary translation approach is the lower complexity of encoding and decoding look-up tables. The disadvantage is, as we show, the loss in information capacity of the binary versus the quaternary approach.

We start in Section~\ref{secGCbal} with a description of the limiting properties of AT/GC-balanced codes, while Section~\ref{secbal} presents code designs for efficiently generating AT/GC-balanced strands. Limiting properties  and code constructions that impose a maximum homopolymer run are discussed in Section~\ref{secmaxrl}. In Section~\ref{seccombbal}, we enumerate the number of binary and quaternary sequences with combined weight and run-length constraints. We specifically compute and compare the information capacity of binary versus `direct' quaternary coding techniques. Section~\ref{conclus} concludes the paper.
\section{AT/GC content balance}\label{secGCbal}
\begin{figure}[t]
\centering
\includegraphics[height=1.9in,width=2.5in]{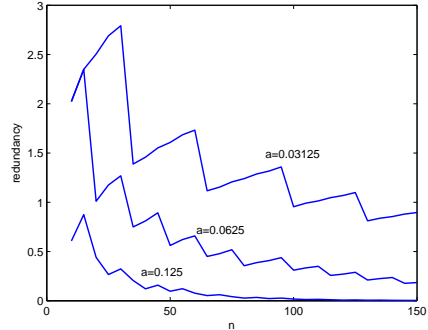}
\caption{\small Redundancy (bits) versus word length, $n$, with the relative unbalance, $a$, as a parameter. The raggedness of the curves is caused by the truncation effects in the summation in (\ref{eqnan1}).}
\label{figunb7}
\end{figure}

We use the nucleotide alphabet ${\cal Q} = \{0,1,2,3\}$, where we propose the following relation between the four decimal symbols and the nucleotides: $G=0, C=1, A=2$, and $T=3$. The {\em AT/GC~content} constraint stipulates that around half of the nucleotides should be either an A or a T~nucleotide. In order to study AT-balanced nucleotides, we start with a few definitions. We define the {\em weight} or {\em AT-content}, denoted by $w_4(\bfx)$, of the $n$-nucleotide oligo $\bfx$ = $(x_1, \ldots, x_n)$, $x_i \in{\cal Q}$, as the number of occurrences of A or T, or
\beq w_4(\bfx) =  \sum_{i=1}^n \varphi(x_i) ,\eeq
where
\beq \varphi(u)= \left\{ \begin{array}{ll}0, & u<2,\\ 1, &  u>1. \end{array}\right. \label{eqw2a}\eeq
The {\em relative unbalance} of a word, $\alpha(\bfx)$, is defined by $\alpha(\bfx) = \left |\frac{w_4(\bfx)}{n} - \frac{1}{2} \right|$. An $n$-nucleotide oligo is said to be balanced if $\alpha(\bfx)=0$. In case we have a set ${\cal S}$ of $n$-symbol codewords, we define the worst case relative unbalance of ${\cal S}$, denoted by $\alpha_{\cal S}$, by $\alpha_{\cal S} = \max_{\bfx \in {\cal S}} \alpha(\bfx)$. Similarly the weight of a binary word $\bfx=$ $(x_1, \ldots, x_n)$, $x_i \in \{0,1\}$, denoted by $w_2(\bfx)$, is defined by
\beq w_2(\bfx) = \sum_{i=1}^n \varphi(2x_i) = \sum_{i=1}^n x_i . \label{eqw2b}\eeq
If we write the 4-ary word $\bfx$ = $(x_1, \ldots, x_n)$, $x_i \in{\cal Q}$, as $\bfx=\bfy+2\bfz$, where both $y_i$ and $z_i \in \{0,1\}$ then
\beq w_4(\bfx) = \sum_{i=1}^n \varphi(x_i) = \sum_{i=1}^n \varphi(2z_i)= w_2(\bfz)  \label{eqw2c} . \eeq
For DNA-based storage, we do not require that the strands of the codebook, ${\cal S}$, are strictly balanced, as a small unbalance, that is $\alpha_{\cal S} \ll 1$, between the GC and AT content is permitted without affecting the error performance. Such a constraint is called a {\em weak balance constraint}. Let ${\cal S}_w$ denote the set of 4-ary words of length $n$ with balance $w=w_4(\bfx)$, or
\beq {\cal S}_w =  \{\bfx \in {\cal Q}^n: w=w_4(\bfx) \} .\eeq
The cardinality of ${\cal S}_w$, denoted by $N(w,n)$, equals
\beq N(w,n) = |{\cal S}_w| = {n \choose w} 2^n. \eeq
The number of oligo's, $N_a(n)$, of length $n$, whose relative unbalance $\alpha(\bfx)\leq a,$ is given by
\beq N_a(n) = \sum_{|w/n-\frac{1}{2}|<a}  N(w,n) = 2^n \sum_{|w/n-\frac{1}{2}|<a} {n \choose w} . \label{eqnan1}\eeq
The redundancy of nearly balanced strands, denoted by $r(a,n)$, equals
\beq  r(a,n) = \log_2 \frac{4^n} {N_a(n)} .\eeq
Figure~\ref{figunb7} shows examples of computations of the redundancy versus $n$ with the relative unbalance, $a$, as a parameter. The raggedness of the curves is caused by the truncation effects in the summation in (\ref{eqnan1}). The distribution for asymptotically large $n$ of $N(w,n)$ versus $w$ is approximately Gaussian shaped, that is
\beq N(w,n)  \sim  {\cal G} \left(w;\frac{n}{2},\frac{n}{4} \right) 4^n, \,\, n \gg 1, \eeq
where
\beq {\cal G}(u;\mu,\sigma^2) = \frac{1}{\sigma\sqrt {2\pi}} e^{-\frac{1}{2} (\frac{u-\mu}{\sigma})^2 } , \eeq
denotes the Gaussian distribution and $\mu$ and $\sigma^2$ denote the mean and variance of the distribution. The number of oligo's, $N_a(n)$, of length $n$, whose relative unbalance $\alpha(\bfx)\leq a,$ is given by~[\cite{Erl}, supplement]
\beq N_a(n)  \sim 4^n \left [1- 2Q (2a \sqrt{n}) \right], \,\, n \gg 1 ,  \eeq
where the $Q$-function is defined by
\beq Q(x) = \frac{1}{\sqrt {2 \pi}} \int_x^\infty e^{-\frac{u^2}{2} } du .\eeq
%
%

In the next section, we discuss various embodiments of codes that balance strands of nucleotides.
\section{Implementations of balanced GC/AT content}\label{secbal} There is a wealth of prior art binary balanced codes~\cite{I54}, and application of such prior art codes to the problem at hand is shown below. Earlier embodiments can be found in~\cite{Tar},~\cite{Yaz}.
\subsection{Binary sequences, Construction~I}
We assume the encoder receives a string of $\ell+n$, $n \geq \ell$, binary symbols, which are translated into a balanced word of $n$ 4-ary symbols. To that end, let $(y_1, \ldots, y_{\ell+n})$, $y_i \in \{0,1\}$, $n\geq \ell$, be an $(\ell+n)$-bit source string. We translate the first $\ell$ bits of the binary source data, $(y_1, \ldots, y_{\ell})$, into a (nearly) balanced binary string $(u_1, \ldots , u_n)$, $u_i \in \{0,1\}$. We merge the $n$-bit string, $(u_1, \ldots, u_n)$, and the remaining $n$-bit segment of the source string, $(y_{\ell+1}, \ldots , y_{\ell+n})$, into the 4-ary vector $\bfv$, $v_i \in {\cal Q}$, using the operation $v_i = y_{\ell+i}+2u_i$, $1 \leq i \leq n$. The balance of the output string, $\bfv$, is given by, see (\ref{eqw2c}), $w_4(\bfv) = w_2(\bfu).$ The rate of the above 4-ary code construction equals $R = 1+\frac{\ell}{n}$. Implementations of balanced codes can be found in the literature. For example, the 8B10B is a binary code of rate 8/10 that has found application in both transmission and data storage systems~\cite{Wi1}. The 10-bit codewords may have four, five or six `one's, and the two-state code guarantees that the unbalance of the encoded sequence is at most $\pm 1$. In case we translate $p$ 8-bit words into $p$ 10-bits words, we have $\alpha_{\cal S}=\frac{1}{10p}$. The (overall) rate $R=\frac{9}{5}$.
\subsubsection{Weak Knuth code}\label{secweakknuth}
Knuth~\cite{Knu} presented an encoding technique for generating binary balanced codewords capable of handling (very) large binary blocks. An $n$-bit user word, $n$ even, is forwarded to the encoder, which inverts the first $k_0$ bits of the user word, where $k_0$ is chosen in such a way that the modified word has equal numbers of ones and zeros. Knuth showed that such an index $k_0$ can always be found. The index $k_0$ is represented by a (preferably) balanced word, called {\it prefix}, of length $p_0$, $p_0 \geq \log_2 n$ bits, so that the redundancy of Knuth's method is approximately $\log_2 n$ (bit). The (balanced) $p_0$-bit prefix and the balanced $n$-bit user word are both transmitted. The receiver can easily undo the inversion of the first $k_0$ bits of the received word. Modifications of the generic Knuth scheme have been presented by Weber~\& Immink~\cite{We3}.

DNA-based storage does not require exact strand GC/AT-content balance, and we may attempt to construct less redundant nearly-balanced codes. We modify Knuth's algorithm for generating nearly balanced binary codes. Let $\bfx=$ $(x_1, \ldots, x_n)$, be the word to be balanced. Define the $m_0=2^{p_0}$ {\em balancing positions}, denoted by $b_i, i=0, \ldots, m_0-1$, that are evenly distributed over the $n$ possible positions, say $b_i=1+is$, $i=0, \ldots, m_0-1$, where $s=\lceil n/m_0 \rceil$. Mimicking the original Knuth encoder, the encoder successively inverts the symbols of the $i$th segment of $\bfx$, $i=0, \cdots, m_0-1$, thereby successively inverting the symbols $x_1$ till $x_{b_0}$, $x_1$ till $x_{b_1}$, etc, until $x_1$ till $x_{b_{m_0-1}}$. The encoder selects the index, $b_{\hat i}$, that enables the least unbalance. In similar vein as in Knuth's method, the index $\hat i$ is represented by a redundant (balanced or nearly balanced) $p$-bit prefix that is appended to the weakly-balanced word. According to Knuth we can choose at least one index $k_0$, $1 \leq k_0 \leq n$, such that exact balance can be achieved. As an `exact' balancing index, $k_0$, is at most $\lfloor s/2 \rfloor$ positions away from position $b_{\hat i}$, we conclude that the relative unbalance is
\beq \alpha_{\cal S} \sim \frac{1}{2^{p_0+1}} . \label{eqafrac} \eeq
The redundancy of the above weak Knuth code equals at least $p_0$ bits (note that additional redundancy is needed to encode the prefix into a nearly balanced word). Let, for example, the code redundancy be $p_0=3$, then $\alpha_{\cal S}=0.0625$. Figure~\ref{figunb7} shows that for a relative unbalance $a=0.0625$ we need, in theory, less than 1.5 bit redundancy for $n>25$, so that we conclude that the above modification of Knuth's algorithm falls far short of the minimum redundancy required. In the next section, we discuss constructions for generating strings that avoid long repetitions of the same nucleotide.
\section{Maximum runlength constraint}\label{secmaxrl}
Long repetitions of the same nucleotide (nt), called a {\em homopolymer run} or {\em runlength}, may significantly increase the chance of sequencing errors~\cite{Bor},~\cite{Ros}, and should be avoided. Avoiding long runs of the same nucleotide will result in loss of information capacity, tand codes are required for translating arbitrary source data into constrained quaternary strings. Binary runlength limited (RLL) codes have found widespread application in digital communication and storage devices since the 1950s~\cite{Cat},~\cite{I54}. MacLaughlin {\em et al.}~\cite{Mc1} studied multi-level runlength limited codes for optical recording. An $n$-nucleotide oligo, a string of 4-ary symbols of length $n$, can be seen as two parallel {\em binary} strings of length $n$, namely a string of a least and a most significant bit with which the 4-ary symbol can be represented. Such a system of multiple parallel data streams with joint constraints is reminiscent of `two-dimensional' track systems, which have been studied by Marcellin and Weber~\cite{Ma7}.

We start in the next subsection with the counting of $q$-ary sequences that satisfy a maximum runlength, followed by subsections where we describe limiting properties and code constructions that avoid $m+1$ repetitions of the same nucleotide.
\subsection{Counting $q$-ary sequences, capacity}
Let the number of $n$-length sequences consisting of $q$-ary symbols have a maximum run, $m$, of the same symbol be denoted by $N_q(m,n)$. The number $N_q(m,n)$ can be found using the next Theorem which defines a recursive relation~\cite{Sh1},~Part~1.
\begin{theorem}\beq N_q(m,n)= \left\{ \begin{array}{ll} q^n,& n \leq m,\\ (q-1)\sum_{k=1}^m N_q(m,n-k),&n>m. \end{array}\right. \label{eqnqmn} \eeq \end{theorem}
{\bf Proof:} For $n \leq m$ the above is trivial as all sequences satisfy the maximum runlength constraint. For $n>m$ we follow Shannon's approach~\cite{Sh1} for the discrete noiseless channel. The runlength of $k$ symbols $a$ can be seen as a 'phrase' $a$ of length $k$. After a phrase $a$ has been emitted, a phrase of symbols $b \neq a$ of length $k$ can be emitted without violating the maximum runlength constraint imposed. The total number of allowed sequences, $N_q(m,n)$, is equal to $(q-1)$ times the sum of the numbers of sequences ending with a phrase of length $k=1,2, \ldots m$, which are equal to $N_q(m,n-k)$. Addition of these numbers yields (\ref{eqnqmn}), which proves the Theorem. \endofproof\\[0.5ex]
Using the above expressions, we may easily compute the feasibility of a $q$-ary $m$-constrained code for relatively small values of $n$ where a coding look-up table is practicable, see Subsection~\ref{secnobin} for more details.

\subsubsection{Generating functions} Generating functions are a very useful tool for enumerating constrained sequences~\cite{Fla}, and they offer tools for approximating the number of constrained sequences for asymptotically large values of the sequence length $n$. The series of numbers $\{N_q(m,n)\}$, $n=1,2 \ldots$, in (\ref{eqnqmn}), can be compactly written as the coefficients of a formal power series $H_{q,m}(x)=\sum N_q(m,i)x^i$, where $x$ is a dummy variable. There is a simple relationship between the generating function, $H_{q,m}(x)$, and the linear homogenous recurrence relation (\ref{eqnqmn}) with constant coefficients that defines the same series ~\cite{Fla}. We first define a generating function
\beq G(x) =  \sum g_{i} x^i . \eeq
Let the operation $[x^{n}] g(x)$ denote the {\it extraction} of the coefficient of $x^n$ in the formal power series $G(x)$, that is, define
\beq [x^{n}] \left (\sum g_{i} x^i\right ) = g_{n}. \eeq
Let
\beq T(x) = \sum_{i=1}^m x^i .\eeq
\begin{theorem} The number of $n$-symbol $m$-constrained $q$-ary words is \beq N_q(m,n) = [x^n]\frac{q T(x)}{1-(q-1)T(x)}. \eeq\end{theorem}
{\bf Proof:} The generating function for the number of $q$-ary sequences with a maximum runlength $m$ is
$$  qT(x) + q(q-1)T(x)^2+ q(q-1)^2T(x)^3 + \cdots .$$
We may rewrite the above as $$\frac{q T(x)}{1-(q-1) T(x)},$$ which proves the Theorem.\endofproof\\[0.5ex]
\subsubsection{Asymptotical behavior}
For asymptotically large codeword length $n$, the maximum number of (binary) user bits that can be stored per $q$-ary symbol, called {\em (information) capacity}, denoted by $C_q(m)$, is given by~\cite{Sh1}
\beq  C_q(m) = \lim_{n \rightarrow \infty} \frac{1}{n} \log_2 N_q(m,n) = \log_2 \lambda_q(m), \eeq
where $\lambda_q(m)$, is the largest real root of the characteristic equation~\cite{Sh1},~\cite{Mc1}
\beq x^{m+1} - q x^m + q-1 = 0  . \eeq
\begin{table}\caption{Capacity $C_2(m)$ and $C_4(m)$ versus $m$.} $$\begin{array}{c|l|l} \hline m & C_2(m) & C_4(m)\\ \hline 1 & 0.0000  &  1.5850 (=\log_2 3) \\2 & 0.6942 & 1.9227 \\3 & 0.8791 & 1.9824\\ 4 & 0.9468 & 1.9957\\5 & 0.9752 & 1.9989\\6 & 0.9881 & 1.9997\\ \hline \end{array}$$ \label{tab1}\end{table}
Table~\ref{tab1} shows the information capacities $C_2(m)$ and $C_4(m)$ versus maximum allowed (homopolymer) run $m$. For asymptotically large $n$ we may approximate $N_q(m,n)$ by~\cite{Fla}
\beq  N_q(m,n) \sim A_q(m) \lambda^n_q (m) . \label{eqNqmn}\eeq
The coefficient $A_q(m)$ is found, see [\cite{I54}, page~157-158], by rewriting $H_{q,m}(x)$ as a quotient of two polynomials, or $H_{q,m}(x)=\frac{r(x)}{p(x)}$. Then
\beq  A_q(m) = - \lambda_q(m)  \frac{r(1/\lambda_q(m) )} {p'(1/\lambda_q(m)) } . \eeq
Table~\ref{tab31} shows the coefficients $A_2(m)$ and $A_4(m)$ versus $m$. For $m=1$, we simply find $N_4(1,n)=4.3^{n-1}$. We found that the approximation (\ref{eqNqmn}) is remarkably accurate. For a typical example, $N_4(2,10)=676836$, while the approximation using (\ref{eqNqmn}) yields $N_4(2,10)\sim 676835.9769$.
\begin{table}\caption{Coefficient $A_2(m)$ and $A_4(m)$ versus $m$.} $$\begin{array}{c|l|l} \hline m & A_2(m) & A_4(m)\\ \hline 1 &        & 1.3333(=4/3) \\2 & 1.4477 & 1.1031 \\3 & 1.2368 & 1.0341\\4 & 1.1327 & 1.0110\\5 & 1.0759 & 1.0034\\6 & 1.0435 & 1.0010\\ \hline \end{array}$$ \label{tab31}\end{table}
The redundancy of a 4-ary string of length $n$ with a maximum runlength $m$, denoted by $r_4(m,n)$, is
\begin{eqnarray} r_4(m,n) &=& 2n-\log_2 N_4(m,n) \nonumber \\ &\sim& n \left(2- C_4(m)\right)-\log_2 A_4(m)  . \end{eqnarray}
\subsection{Binary-based RLL code construction, Construction II}\label{subsecbincode}
In a similar vein as presented in Section~\ref{secbal}, we may exploit binary maximum runlength limited (RLL) codes for generating quaternary RLL sequences. Construction~II exemplifies such a technique for $m>1$. Let  $\bfu = (u_1, \ldots , u_n)$ be an $n$-bit RLL string. We merge the RLL $n$-bit string, $\bfu$, with an $n$-bit source string $\bfy=(y_1, \ldots , y_n)$, by using the addition $v_i = u_i+2y_i$, $1 \leq i \leq n$, where $\bfv=(v_1, \ldots, v_n)$, $v_i \in {\cal Q}$ is the 4-ary output string. It is easily verified that the 4-ary output string, $\bfv$, has maximum allowed run $m$, the same as the binary string $\bfu$. The number of distinct 4-ary sequences, $\bfv$, of Construction~II equals $2^nN_2(m,n)$, so that the redundancy, denoted by $r_2(mn,n)$ is
\beq r_2(m,n)  \sim n \left(1- C_2(m)\right)-\log_2 A_2(m) . \label{eqr2mn}  \eeq
The capacity loss with respect to the runlength limited 4-ary channel, denoted by $\eta(m)$, is expressed by
\beq \eta(m) = \frac{1+C_2(m)}{C_4(m)} . \label{eqetam} \eeq
\begin{table}\caption{Asymptotic rate efficiency, $\eta(m)$, of binary Construction~II versus maximum homopolymer run, $m$.} $$\begin{array}{l|l} \hline m & \eta(m) \\ \hline 2 & 0.881\\3 & 0.948\\4 & 0.975\\5 & 0.988\\ 6 & 0.994\\7 & 0.997\\\hline \end{array}$$ \label{tab611}\end{table}
Table~\ref{tab611} lists results of computations. We may notice that for small values of $m$, Construction~II will suffer a capacity loss of up to 12~\% for $m=2$. For larger values of $m$, however, the capacity loss is negligible.

The above asymptotic efficiency of Construction~II, $\eta(m)$, is valid for very large values of the strand length $n$, and it is of practical interest to assess the efficiency for smaller values of the strand length. Construction~II can be used with any binary RLL code, and there are many binary code constructions for generating maximum runlength constrained sequences, see~\cite{I54} for an overview. We propose here, for the efficiency assessment, a simple two-mode block code of codeword length $n$. Runlength constrained codewords in the first mode start with a symbol `zero', while codewords in the second mode start with a `one'. When the previous sent codeword ends with a `one' we use the codewords from the first mode and {\em vice versa}. The number of binary source words that can be accommodated with Construction~II equals $2^{n-1} N_2(m,n)$, so that the code rate, denoted by $R_{m,0}$, is
\beq R_{m,0} = \frac{1}{n} \left (n-1 + \lfloor \log_2 N_2(m,n) \rfloor \right) , \eeq
where we truncated the code size to the largest power of two possible.
\begin{table}\caption{Rate efficiency, $R_{m,0}/C_4(m)$, of binary Construction~II versus strand length, $n$, and maximum homopolymer run, $m$.} $$\begin{array}{c|l|l|l} \hline n & m=2 & m=3 &m=4 \\ \hline
5  & 0.832 & 0.807 & 0.802\\6  & 0.780 & 0.841 & 0.835\\7  & 0.817 & 0.865 & 0.859\\8  & 0.845 & 0.883 & 0.877\\9  & 0.809 & 0.897 & 0.891\\10 & 0.832 & 0.908 & 0.902\\\hline \end{array}$$ \label{tab411}\end{table}
Table~\ref{tab411} shows selected outcomes of computations of the rate efficiency $R_{m,0}/C_4(m)$ versus $m$ and $n$.
\subsection{Encoding of quaternary sequences without binary step}\label{secnobin}
In this subsection, we investigate constructions of codes that transform binary words directly (that is, without an intermediate binary coding step) into 4-ary maximum homopolymer constrained codewords. An example of a simple 4-ary block code was presented by Blawat {\em et al.}~\cite{Bl7}. The code converts 8~source bits into a 4-ary word of 5~nt. The 5-nt words can be cascaded without violating the prescribed $m=3$ maximum homopolymer run. The rate of Blawat's construction is $R=8/5=1.6$. As $C_4(m=3)=1.9824$, see Table~\ref{tab1}, the (rate) efficiency of the construction is $R/C_4(m)=0.807$. Alternative, and more efficient, constructions are described below.
\subsubsection{State-independent decoding}
A source word can be represented by two $n$-symbol 4-ary $m$-constrained codewords. The two representations  differ at the first position. In case we cascade a new codeword to the previous codeword, we are always able to choose (at least) one representation whose first symbol differs from the last symbol of the previous codeword. Then, clearly, the cascaded string of 4-ary symbols satisfies the maximum homopolymer run constraint. The rate of this two-mode construction, denoted by $R_{m,1}$, is
\beq R_{m,1} = \frac{1}{n} (\lfloor \log_2(N_4(m,n))\rfloor-1), \eeq
where we truncated the code size to the largest power of two possible. Table~\ref{tab311} shows selected outcomes of computations of the rate efficiency $R_{m,1}/C_4(m)$ versus $m$ and $n$. We observe that, for $m=2$, the 'quaternary' efficiency $R_{2,1}/C_4(2)$ is slightly better than the 'binary' $R_{2,0}/C_4(2)$, For $m>2$, both approaches have the same efficiency.
\begin{table}\caption{Rate efficiency, $R_{m,1}/C_4(m)$, of the 4-ary code construction versus strand length, $n$, and maximum homopolymer run, $m$.} $$\begin{array}{c|l|l|l|l} \hline n & m=1 & m=2 & m=3 &m=4 \\ \hline
5 & 0.883 & 0.832 &  0.807 &  0.802 \\ 6 & 0.841 & 0.867 & 0.841 & 0.835 \\
7 & 0.901 & 0.892 &  0.865 &  0.859 \\ 8 & 0.946 & 0.910 & 0.883 & 0.877 \\
9 & 0.911 & 0.925 &  0.897 &  0.891 \\10 & 0.946 & 0.936 &  0.908 &  0.902 \\ \hline \end{array}$$ \label{tab311}\end{table}
The conversion of the binary source symbols into the 4-ary $n$-nt strands and vice versa can be accomplished using look-up tables of complexity $4^n$.
\subsubsection{State-dependent decoding}
In the above construction, the encoded codeword depends on the last symbol of the previous codeword. Decoding, however, is based on the observation of the $n$ symbols of the retrieved codeword. In this subsection, we discuss a state-dependent decoding construction, where the codeword chosen depends on the last symbol of the previous codeword, and decoding is based on the observation of the $n$ symbols of the retrieved codeword plus the last symbol of the previous codeword. We define four tables of codewords, denoted by $L(i,a)$, where $i$, $1\leq i \leq K$, denotes the decimal representation of the source word to be encoded, $K$ denotes the size of the table, and $a$ denotes the encoder state $a=\in \{1,2,3,4\}$. We construct the four tables in such as way that the codewords in each table $L(i,a)$ do not start with the symbol $a$. Then, the maximum size of the tables equals $K=\frac{3}{4} N_4(m,n)$ (note that $N_4(m,n)$ is a multiple of 4). The  representation, $L(i,a)$, chosen depends on the last symbol of the previous codeword, $a$. The rate of this four-mode construction, denoted by $R_{m,2}$, is
\beq R_{m,2} = \frac{1}{n} \left \lfloor \log_2 \left(\frac {3}{4} N_4(m,n) \right) \right\rfloor . \eeq
Table~\ref{tab312} shows the rate efficiencies that can be reached with this construction. The efficiency improvement with respect to Table~\ref{tab311} is obtained at the cost of a four times larger look-up table. Decoding of a codewords is uniquely accomplished by observing the $n$-symbol codeword plus the last symbol of the previous codeword.
\begin{table}\caption{Rate efficiency, $R_{m,2}/C_4(m)$, of the 4-ary code construction versus strand length, $n$, and maximum homopolymer run, $m$.} $$\begin{array}{c|l|l|l|l} \hline n & m=1 & m=2 & m=3 &m=4 \\ \hline
5 & 0.883 & 0.936 & 0.908 & 0.902 \\  6 & 0.946 & 0.954 & 0.925 & 0.919  \\
7 & 0.991 & 0.966 & 0.937 & 0.931 \\  8 & 0.946 & 0.975 & 0.946 & 0.940  \\
9 & 0.981 & 0.982 & 0.953 & 0.946 \\ 10 & 0.946 & 0.936 & 0.958 & 0.952 \\\hline \end{array}$$ \label{tab312}\end{table}
\begin{example} Let (as in Blawat's code~\cite{Bl7}) $n=5$ and $m=3$. We simply find, using (\ref{eqnqmn}), $N_4(3,5)=996$, so that the code may accommodate $K=3/4 \times 996=747$ binary source words. Since $K>512=2^9$ we may implement a code of rate 9/5, which is 12\% higher than that of Blawat's code of rate 8/5. As we have the freedom of deleting 747-512=235 redundant codewords, we may bar the words with the highest unbalance. \endofproof\\\end{example}
In the next section, we take a look at the combination of balance and maximum polymer run constrained codes.
\section{Combined weight and maximum run constrained codes}\label{seccombbal}
Kerpez {\em et al.}~\cite{Ke1}, Braun and Immink~\cite{Br3}, and Kurmaev~\cite{Kur} analyzed properties and constructions of binary combined weight and runlength constrained codes. Their results are straightforwardly applied to the quaternary case at hand. In the next section, we count binary and quaternary sequences that satisfy combined maximum runlength and weight constraints. We start by counting the number of binary sequences, $\bfx$, of length $n$ that satisfy a maximum runlength constraint $m$ and have a weight $w=w_2(\bfx)$. Paluncic and Maharaj~\cite{Pal} enumerated this number for the balanced case $w=w_2(\bfx)=0$.
\subsection{Counting binary RLL sequences of given weight}\label{seccountbin}
Define the bi-variate generating function $H(x,y)$ in the dummy variables $x$ and $y$ by

\beq H(x,y) = \sum_{i,j} h_{i,j} x^i y^j, \eeq
and let $[x^{n_1}y^{n_2}] h(x,y)$ denote the extraction of the coefficient of $x^{n_1}y^{n_2}$ in the formal power series $\sum h_{i,j} x^i y^j$, or
\beq [x^{n_1}y^{n_2}] \left (\sum h_{i,j} x^i y^j \right ) = h_{n_1,n_2}. \eeq
Define
\beq T_1(x,y) = \sum_{i=1}^m x^i y^i. \eeq
The number of $n$-bit codewords, $\bfx$, with maximum runlength $m$, denoted (with a slight abuse of notational convention by adding an extra parameter) by $N_2(m,w,n)$, that satisfy a given unbalance constraint $w=w_2(\bfx)$ is given by the next Theorem.
\begin{theorem}
$$ N_2(m,w,n) = [x^n y^w] \frac{T_1(x,y)+T(x)+2T_1(x,y)T(x)}{1-T_1(x,y)T(x)} . $$
\end{theorem}
{\bf Proof:} Let the sequence start with a runlength of zero's, then the generating function for the number of binary sequences with a maximum runlength $m$ is
$$  T(x) + T(x)T_1(x,y) + T(x)^2T_1(x,y)+ T(x)^2T_1(x,y)^2 + \cdots .$$
In case the sequence starts with a run of one's, we obtain for the generating function
$$ T_1(x) + T(x)T_1(x,y) + T(x)T_1(x,y)^2+ T(x)^2T_1(x,y)^2 + \cdots . $$
The generating function for the number of binary sequences with a maximum runlength $m$ starting with a one or a zero runlength is the sum of the two above generating functions. Working out the sum yields
$$ \frac{T_1(x,y)+T(x)+2T_1(x,y)T(x)}{1-T_1(x,y)T(x)} , $$
which proves the Theorem.\endofproof\\[0.5ex]
With the above bi-variate generating function, we may exactly compute the number of binary $m$-constrained words of weight $w$. More insight is gained by an approximation of $N_2(m,w,n)$. For a given maximum runlength, $m$, and large $n$, we are specifically interested in the distribution of $N_2(m,w,n)$ versus the weight $w$. For asymptotically large $n$, according to the central limit theorem, the distribution of the number of sequences versus weight, $w$, is approximately Gaussian~\cite{Fla}.
\begin{theorem} \beq N_2(m,w,n) \sim {\cal G} \left(w;\frac{n}{2}, \frac{\gamma_2(m)n}{4} \right) N_2(m,n) ,\label{eqn2mwn}\eeq
where \beq \gamma_2(m)   = \frac{1}{\bar l} \sum_{i=1}^m (i-\bar l)^2 \lambda_2^{-i}(m) \eeq
and
\beq \bar l = \sum_{i=1}^m i \lambda_2^{-i}(m) . \eeq \end{theorem}
{\bf Proof:} The probability of occurrence of a runlength of length $k$, $k \leq m$, is $\lambda_2^{-k}(m)$, see~\cite{I54}, Chapter 4. So that the average number of runlengths in a sequence of $n$ symbols is $n/{\bar l}$. The weight $w$ is the sum of the runlengths of ones, so that according to the central limit theorem the weight distribution is approximately Gaussian for large $n$ with mean $\frac{n}{2}$ and variance $\frac{\gamma_2(m)n}{4}$.  \endofproof\\[0.5ex]

Table~\ref{tabdist2} shows results of computations (the parameter $\gamma_4(m$) is explained in Section~\ref{counting4ary}). Perusal of the outcomes clearly demonstrates that for small values of $m$ the unbalance variance, $\gamma_2(m) n$, is smaller than that of unconstrained sequences (that is, $m=\infty$) of the same length $n$. In other words, a maximum runlength `helps' to reduce the expected unbalance.
\begin{table}\caption{Coefficient $\gamma_2(m)$ and $\gamma_4(m)$ versus maximum homopolymer run $m$.} $$\begin{array}{r|ll} \hline m & \gamma_2(m) & \gamma_4(m)\\ \hline 1 &   & 0.5000\\ 2&  0.1708 & 0.7410  \\ 3&  0.3449 & 0.8796\\  4&  0.5059& 0.9497 \\ 5&  0.6426 & 0.9808\\ 10& 0.9565 & 0.9999  \\ \infty & 1 & 1 \\ \hline \end{array}$$ \label{tabdist2}\end{table}
\subsection{Counting quaternary RLL sequences of given weight}\label{counting4ary}
We count the number of $n$-tuples $\bfx$ of 4-ary symbols that satisfy a maximum run length constraint, $m$, and have weight $w=w_4(\bfx)$, denoted (with a slight abuse of notational convention) by $N_4(m,w,n)$.
\subsubsection{Maximum runlength constraint}
For the special case $m=1$, Limbachiya~\cite{Lim} {\em et al.} presented a closed expression of $N_4(1,w,n)$. For other values of the prescribed maximum runlength, $m$, we may readily compute the number of 4-ary sequences, $N_4(m,w,n)$, versus weight, $w=w_4(\bfx)$, by applying generating functions. The 4-ary symbols are generated by a constrained data source that can be modelled as a four-state Moore-type finite-state machine. The machine steps from state to state where when state $i \in{\cal Q}$ is visited a sequence of $k$, $ 1 \leq k \leq m$, symbols `$i$' are emitted. After visiting state $i$, the data source may not return to state $i$ (and thus emit a sequence of the same symbol `$i$' again), but it enters state $j \neq i$, $j \in{\cal Q}$. When the machine enters state~3 or 4, the word weight, $w$, is incremented by $k$, where $k$, $ 1 \leq k \leq m$, denotes the run of symbols `3' or `4'. When, on the other hand, states~1 or 2 are entered, the weight increment is nil. The resulting $4 \times 4$ one-step {\em skeleton} or {\em state-transition} matrix, $D(x,y)$, of the finite-state machine is
\beq D(x,y)=\left [\begin{array}{cccc}  0 & a_0 & a_0 & a_0 \\ a_0 & 0 & a_0 & a_0\\a_1 & a_1 & 0 & a_1 \\ a_1 & a_1 &a_1 & 0\\\end{array}\right], \eeq
where $a_0=T(x)$ and $a_1=T_1(x,y)$.
\begin{theorem} {\em The number of 4-ary sequences of length $n$ with maximum runlength constraint $m$ and weight $w$ equals
\beq N_4(m,w,n) = [x^n y^w] \frac{1}{3} \sum_{i,j} d_{i,j}^{[n]}(x,y), \label{eqn4mwn}\eeq
where $d_{i,j}^{[n]}(x,y)$ denotes the entries of $D^n(x,y)$.}\end{theorem} {\bf Proof:} The entries $d_{i,j}^{[n]}(x,y)$ of $D^n(x,y)$ are equal to the number of sequences (paths) of length $n$ starting in state $i$ and ending in state $j$. Summation of the entries and division by 3 yields the generating function of $N_4(m,w,n)$.\endofproof\\[0.5ex]

In the next subsection, we derive a simple approximation to $N_4(m,w,n)$ valid for large $n$.
\subsubsection{Estimate of the weight distribution}
For asymptotically large $n$, the weight distribution is approximately Gaussian, that is, we may conveniently approximate $N_4(m,w,n)$ using the next Theorem.
\begin{theorem} \beq N_4(m,w,n) \sim {\cal G} \left(w; \frac{n}{2}, \sigma^2_4(m,n) \right) N_4(m,n), \,\, n\gg 1,\label{eqN4approx} \eeq
where $\sigma^2_4(m,n)$, denotes the variance of the Gaussian weight distribution.
\end{theorem} {\bf Proof:} Let $u_i$, $i=1,2, \ldots$, $u_i \in {\cal Q}$, be an infinitely long 4-ary sequence generated by a maxentropic source that satisfies a prescribed maximum runlength, $m$. Note that although the 4-ary sequence $u_i$, $i=1,2, \ldots$, satisfies a limited runlength constraint, $m$, that runs of the binary weight sequence $v_i=\varphi(u_i)$, $i=1,2, \ldots$, are without limit. The variance, $\sigma^2_4(m,n)$, of the Gaussian weight distribution is governed by the runlength distribution, $P(k)$, of the binary sequence $v_i$, where $P(k)$, $k>0$, denotes the probability of occurrence of a runlength $k$. Clearly, $\sum_{k>0} P(k)=1$. The probability $P(k)$ is given by
\beq  P(k) = c N_2(m,k) \lambda_4^{-k} , \,\,  k \geq 1,\eeq
where the normalization constant $c$ is chosen such that $\sum_{k=1}^{\infty} P(k)=1$. The term $N_2(m,k)$ is the number of AT combinations of length $k$, which may exist of a single A or T run or a plurality of alternating A~and T~runs. We have $\sigma^2_4(m,n)=\frac{\gamma_4(m)n}{4}$, where, see~\cite{I54}, Chapter~4,
\beq \gamma_4(m) = \frac{1}{\bar l} \sum_{k=1}^{\infty} (k-\bar l)^2 P(k) \eeq
and \beq \bar l = \sum_{k=1}^{\infty} k P(k) . \eeq \endofproof

Table~\ref{tabdist2} shows results of computations of $\gamma_4(m)$ versus $m$. We may notice that the weights of the quaternary RLL sequences are more concentrated around the mean $n/2$ than those of binary RLL sequences. The above outcome is not consistent with the results by Ehrlich and Zielinski~\cite{Erl}, as they assume that the balance variance equals $n/4$, independent of $m$.
\subsection{Redundancy of binary and quaternary codes with combined constraints}
As in Constructions~I and~II, let the quaternary word $\bfx = (x_1, \ldots, x_n)$, $x_i \in {\cal Q}$, be written as $\bfx=\bfy+2\bfz$, where the constituting elements $y_i$ and $z_i \in \{0,1\}$. If the binary sequence $\bfz$ is $m$-constrained and has a weight $w=w_2(\bfz)$, then $\bfx$ is $m$-constrained and it has a weight $w_4(\bfz)=w$. The redundancy of the binary constrained sequences, $\bfz$, denoted (with a slight abuse of convention) by $r_2(m,a,n)$, equals
\beq r_2(m,a,n) = n - \log_2  N_2(m,w,n) . \eeq
Using (\ref{eqr2mn}) and (\ref{eqn2mwn}), we obtain for $n \gg 1$, that the redundancy of the binary approach is
\beq r_2(m,a,n) \sim r_2(m,n) - \log_2 \left[1-2Q \left(2a\sqrt{\frac{n}{\gamma_2(m)}}\right )\right]  . \eeq
The redundancy of the quaternary approach, denoted by $r_4(m,a,n)$, equals, for $n \gg1$,
\begin{eqnarray} r_4(m,a,n) &=& \log_2 \frac{4^n}{N_4(m,w,n)}  \\
&\sim& r_4(m,n) - \log_2 \left[1-2Q \left(2a\sqrt{\frac{n}{\gamma_4(m)}} \right) \right] . \nonumber\end{eqnarray}
A numerical analysis of the above expressions shows that the redundancy difference due to the balance (right hand) term is around 0.5-1 bit for $m=2$. For larger values of the homopolymer run $m$ the extra redundancy is negligible. The redundancy difference, $r_2(m,n)-r_4(m,n)$, due to the imposed runlength constraint is much larger for $n>10$ than the redundancy due the balance constraint. For $m>6$ the difference between $r_2(m,n)$ and $r_4(m,n)$ is negligible, see Subsection \ref{subsecbincode}, so that considering the much larger look-up tables needed for quaternary codes, the binary approach using Construction~1 for combined constraints is preferable from a practical point of view.
\section{Conclusions}\label{conclus}
We have described coding techniques for weakly balancing GC and AT-content and avoiding homopolymer runs larger than $m$ nt's of quaternary DNA strings. We have found exact and approximate expressions for the number of binary and quaternary sequences with combined weight and run-length constraints. We have compared two coding approaches for constraint-based coding of DNA strings. In the first approach, an intermediate, `binary', coding step is used, while in the second approach we `directly' translate  source data into constrained quaternary sequences. The binary approach is attractive as it yields a lower complexity of encoding and decoding look-up tables. The redundancy of the binary approach is higher than that of the quaternary approach for generating combined weight and run-length constrained sequences. The redundancy difference is small for larger values of the maximum homopolymer run.
\end{document}